\documentstyle[12pt]{article}
\begin{document}

\begin{center}
{\large "GAUGING" THE FLUID }\\
\vskip 2cm
Subir Ghosh\\
\vskip 1cm
Physics and Applied Mathematics Unit,\\
Indian Statistical Institute,\\
203 B. T. Road, Calcutta 700108, \\
India.
\end{center}
\vskip 3cm
{\bf Abstract:}\\
A consistent framework has been put forward to quantize the
isentropic, compressible and inviscid fluid model in the
Hamiltonian framework, using the Clebsch parameterization. The
naive quantization is hampered by the non-canonical (in particular
field dependent) Poisson Bracket algebra. To overcome this
problem, the Batalin-Tyutin \cite{12} quantization formalism is
adopted in which the original system is converted to a local gauge
theory and is embedded in a {\it canonical} extended phase space.
In a different reduced phase space scheme \cite{vy} also the
original model is converted to a gauge theory and subsequently the
two distinct gauge invariant formulations of the fluid model are
related explicitly. This strengthens the equivalence between the
relativistic membrane (where a gauge invariance is manifest) and
the fluid (where the gauge symmetry is hidden).
Relativistic
generalizations of the extended model is also touched upon.

\vskip 3cm
\noindent
PACS Numbers: 11.15.-q, 47.65.+a   \\
Key Words: gauge theory, BT quantization,  fluid.

\newpage
\section{Introduction}
\label{sec:Introduction}
Extended objects, a generic example being $d$-branes \cite{br}, are receiving more and more
attention in high energy physics. The equivalence between membrane ($d=2$ brane) and
planar fluid mechanics was revealed some years ago \cite{br1}. Subsequently this connection
was
established in arbitrary dimensions \cite{1}. In fact a specific fluid model,
 the Chaplygin gas in $(d,1)$ space-time and the Poincare invariant Born-Infeld
model in $(d,1)$ space-time, both bear a common ancestry to the
reparameterization invariant Nambu-Goto $d$-brane theory in
$(d+1,1)$ space-time. These issues are discussed in \cite{1,2}.

The concept of symmetry and in particular local gauge invariance
has permeated through the present day theoretical physics. In
fact, the gauge invariance, (specifically the reparameterization
invariance), of the Nambu-Goto theory plays a key role in
establishing the above mentioned equivalence, since a choice of
light-cone gauge or parameterisation in the Nambu-Goto action
leads to the Chaplygin gas, whereas a Cartesian parameterisation
or a combination of the light-cone gauge and a hodographic
transformation, (in which the independent and dependent variables
are interchanged), yields the Born-Infeld model \cite{1}. These
mappings clearly establish the complete integrability of the $d=1$
Chaplygin gas \cite{lan} and $d=1$ Born-Infeld model \cite{bar}
since the parent Nambu-Goto $1$-brane (string) moving in a
$2$-space (plane) is completely integrable.

The Hamiltonian formulation of the inviscid fluid was initiated by the work of Landau \cite{landau}, who provided the Hamiltonian and the Poisson Bracket algebra of the degrees of freedom, (density and velocity fields), comprising the fluid system.
The discussion concerning the Chaplygin gas model pertains
to {\it irrotational} fluid. (Obviously there can not be any
vorticity in one spatial dimension.) The Hamiltonian system for
irrotational fluid (in any dimension) becomes very simple , with
the velocity field expressible as a gradient of a single scalar
degree of freedom (Clebsch variable \cite{5}). However, it has
been pointed out in \cite{wot} that treated as constraint system
in the Dirac Hamiltonian framework \cite{10}, the irrotational
fluid does not show the gauge invariance, which manifests itself
as a $U(1)$ phase invariance in the relativistic membrane. A
Batalin-Tyutin (BT) extended space analysis \cite{12} of the above
model \cite{10a} restores the gauge invariance, albeit in the
extended phase space. Subsequently, a connection was established
in \cite{wot} between the (extended space) gauge invariance
\cite{10a} and a gauge invariance in the physical sector
\cite{wot}. The latter result exploits the reduced space scheme
discussed in \cite{vy}. In this sense the gauge invariance is hidden \cite{wot} in the
Hamiltonian fluid model.

Let us now put our work in the proper perspective. We do not
restrict ourselves to irrotational fluid and instead concentrate
on the the general case where the fluid can possess vorticity as
well. Essentially, in the next three sections, we have carried out
a similar analysis as that of \cite{10a,wot} for {\it arbitrary}
fluid motion. There are qualitative changes since the velocity
becomes a {\it non-linear} combination three Clebsch variables
\cite{3,4}. This non-linearity makes the BT \cite{12} analysis and
reduced space analysis \cite{vy} quite involved. Interestingly, we
have shown that in the general case of arbitrary fluid motion as
well, there exists a  mapping between the gauge symmetries in the
extended space approach \cite{12} and reduced space scheme
\cite{vy}. However, the simple $U(1)$ gauge group of the
irrotational case is now replaced by the symmetry group of
orientation and area preserving diffeomorphism \cite{st}.

 An earlier work in the context of quantization of fluid system with vortices is the classic paper by Rasetti and Regge \cite{new0} where the fluid under consideration is incompressible that is of constant density. Since we are dealing with a compressible fluid, results equivalent to that of \cite{new0} can be recovered by further constraining our system to a constant density one. We will return to further comparisons between our formalism and that of \cite{new0} in Section {\bf 3}.

The study of the non-relativistic, isentropic, inviscid fluid has emerged lately as an
area of intense activity, principally due to its broad applicability. The solutions of
the Galilleo invariant system in $d$-dimensions offer solutions for the $d+1$-dimensional
relativistic membrane \cite{baz}. Also, the subject has identifications with the hydrodynamical
description of quantum mechanics \cite{mad}, parton model \cite{jev}, hydrodynamics of
superfluids \cite{sh} and black hole cosmology \cite{kam}.

The paper is organized as follows: In section {\bf 2}, we
reproduce briefly the conventional Hamiltonian fluid dynamics.
Subtleties regarding the introduction of the Clebsch variables are
also mentioned. Sections {\bf 3} and {\bf 4} deal with the
construction of a local gauge theory for the fluid. Specifically,
section {\bf 3} consists of the BT extension scheme and {\bf 4}
constitutes the reduced space formalism. Relativistic
generalizations of the above modified fluid models are also
touched up on. The paper ends with a conclusion in {\bf 5}.
\section{Hamiltonian description of the fluid}
\label{sec:HamiltonianDescriptionOfTheFluid}
 Let us start by providing the conventional Hamiltonian formulation of fluid dynamics \cite{1}.
 We will be considering inviscid,
isentropic and compressible fluid in three space dimensions, whose dynamics is governed by
the continuity equation and Euler equation,
\begin{equation}
\partial _t \rho +\partial _i(\rho v^i)=0~,~~
\partial _t v_i+(v^j\partial _j)v_i=f_i,
\label{1}
\end{equation}
where $\rho $ and $v_i$ denote the density and velocity fields
respectively.
We keep $f_i$ arbitrary for the time being.
 The above equations of motion
(\ref{1}) are generated by
 \begin{equation}
 \partial _t\rho (x)=\{\rho (x),H\}~,\partial _tv_i (x)=\{v_i(x),H\}~,
 ~~H[\rho ,v_i]=\int d^3y {\cal H}(y),
  \label{3}
 \end{equation}
using the following
Hamiltonian and the (non-canonical) Poisson Bracket (PB) algebra \cite{landau}
 $$
 {\cal H}={1\over 2}\rho v_iv_i +U~,$$
 \begin{equation}
 \{\rho (x),\rho (y)\}=0~,~~
 \{v_i(x),\rho (y)\}={{\partial }
 \over {\partial x^i}}\delta (x-y)~,~~\{v_i(x),v_j(y)\}=
 -{{\partial _iv_j-\partial _jv_i}\over {\rho }}\delta (x-y).
 \label{4}
 \end{equation}
Note that we consider only those $f_i$ in (\ref{1}) which can be
generated by some $U$. 

It is worth mentioning that generalizations of the $\{v_i(x),v_j(y)\}$ PB in (\ref{4}) has been discussed in the literature \cite{new1}, where $\rho $ in the denominator is replaced by $(\rho -C_0)$, $C_0$ being a dynamical invariant of the system. Different limiting values for $C_0$ lead to distinct regimes of physical interest. The present system with $C_0=0$ was derived before for superfluid in zero temperature limit.

However, a canonical Lagrangian formulation of the above is lacking
due to the presence of the fluid helicity term,
$$C=\int d^3x(\epsilon _{ijk}v^i\partial _jv^k),$$
which, being a Casimir of the theory, creates an obstruction in
the inversion of the symplectic matrix \cite{4,1}. To overcome
this problem, albeit in the special case of helicity-less flows,
Clebsch parameterization
 \cite{5}
of the velocity field $v_i$ and its associated PB algebra are introduced,
\begin{equation}
 v_i(x)\equiv\partial _i\theta (x) +\alpha (x)\partial _i\beta (x),
 \label{cv}
\end{equation}
 \begin{equation}
\{ \theta (x),\alpha (y)\}=-{\alpha \over \rho }\delta (x-y)~,~
 \{\beta (x),\alpha (y)\}={1\over\rho }\delta (x-y)~,~
\{ \theta (x),\rho (y)\}=\delta (x-y).
 \label{6}
 \end{equation}
These are the only non-zero PBs. This parameterization renders the
helicity variable to a surface term without any bulk contribution
and obviously reproduces the previous equations of motion. The
Lagrangian providing the correct symplectic structure \cite{1} and
equations of motion is
\begin{equation}
{\cal L}=
 \dot \theta \rho +\dot \beta
\alpha \rho - ({1\over 2}\rho v_iv_i +U).
\label{g1}
\end{equation}
In our
subsequent discussions, we will always use the Clebsch variables.

The non-canonical algebra, posited in (\ref{4}) or equivalently in (\ref{6}), is field dependent and hence not conducive for
a quantization programme, since the basic Green functions, ({\it i.e.} the two-point functions
or propagators), can not be defined in a perturbative framework. This motivates us to consider
the gauge invariant formulation of the fluid.
\section{Extended space (BT) quantization}
\label{sec:ExtendedSpaceBTQuantization}
We start by digressing a little on the  constraint analysis of Dirac \cite{10} in a
Hamiltonian framework. In this scheme, the constraints are termed
as First Class Constraints (FCC) if they commute (in the PB sense,
 modulo constraints) or Second Class Constraints (SCC)
if they do not. The FCCs induce gauge invariance in the theory
whereas the SCCs tend to modify the symplectic structure of the
phase space for compatibility with the SCCs. The above modification induces a replacement of
the PBs by Dirac Brackets (DB) \cite{10} as defined below,
\begin{equation}
\{A(x),B(y)\}_{DB}= \{A(x),B(y)\}-\int (d^3zd^3w)\{A(x),\eta _\alpha(z)\}
\{\eta _\alpha (z),\eta _\beta (w)\}^{-1}\{\eta _\beta (w),B(y)\}.
\label{dirac}
\end{equation}
where $\eta _\alpha (x)$ refer to th SCCs.

Primarily, we would like to construct a canonical phase space from
which the non-canonical PB algebra  can be derived
naturally as DBs \cite{8}. This requires an embedding of the
original system in a larger phase space having  independent and
commuting canonical pairs,
\begin{equation}
(\theta ,\Pi _\theta \equiv \rho )~~;~~ (\alpha ,\Pi _\alpha )~~;~~
(\beta ,\Pi _\beta ),
\label{can}
\end{equation}
with
  $\{\theta (x),\Pi _\theta (y)\}=\delta (x-y)$ etc. Since we have
introduced two extra variables in
$\Pi _\alpha $ and $\Pi _\beta $,
 we also introduce two SCCs \cite{8}
\begin{equation}
\eta _1\equiv \alpha \Pi_\theta -\Pi _\beta ~;~~\eta _2\equiv \Pi _\alpha .
 \label{10}
 \end{equation}
 In the present case, the SCCs in (\ref{10})
 reproduce (\ref{6}) as DBs from the
above canonical set and also keep the
degrees of freedom count same as the original one.
\footnote {It should be pointed out that the non-canonical algebra (\ref{6}) can be reproduced most economically by exploiting the symplectic quantization formalism \cite{1,new2}. However, we have pursued the Dirac schme \cite{10} here, keeping in mind the subsequent quantization in the BT formalism \cite{12}.}

However, in general, the DB formalism \cite{10} can create
problems in quantizing the theory since the modified symplectic
structure may become field dependent, (if non-linear constraints
are present \cite{11}), making it difficult for them to be elevated to
quantum commutators via the correspondence principle. With the
non-linear constraints (\ref{10}), precisely this situation is
encountered in the present case giving rise to field dependent DBs
(\ref{6}).

To overcome the above mentioned pathology, a systematic procedure has been formulated by
 Batalin and Tyutin (BT) \cite{12} whereby one enlarges the phase
 space, and subsequently the constraints as well as the Hamiltonian
 accordingly, so that the SCCs turn into FCCs in the enlarged
 canonical phase space. Quantization of the resulting gauge theory
 is well understood. Essentially one is trading the original system of physical
 degrees of freedom having a non-canonical PB (or equivalently DB) structure and (possibly)
 a simple looking Hamiltonian with an equivalent, {\it extended}  model having {\it canonical}
 phase space and (possibly) a more complicated looking Hamiltonian. The obvious advantage of
 the latter over the former is that
 the BT extended model lives in a canonical phase space and hence is suitable for
 perturbative quantization. The relevant formulas  regarding the BT extension for our purpose are listed below.

 Let us consider a generic set of constraints $(\Theta _\alpha ,\Psi _l
)$ and a Hamiltonian operator $H$ with the following PB relations,
$$
\{\Theta _\alpha (q) ,\Theta _\beta (q)\}\approx
\Delta ^{\mu\nu}_{\alpha \beta
 }(q) \ne 0~~;~~\{\Theta _\alpha (q) ,\Psi _l (q)\}\approx 0
$$
\begin{equation}
 \{\Psi _l(q) ,\Psi _n (q)\}\approx 0 ~~;~~
\{\Psi _l(q) ,H (q)\}\approx 0.
\label{01}
\end{equation}
In the above $(q)$
collectively refers to the set of variables
present prior to the BT extension
and "$\approx $"
means that the equality holds on the constraint surface. Clearly
$\Theta
_\alpha  $
and $\Psi _l $ are SCC and FCC \cite{10} respectively.

In systems with non-linear SCCs,
(such as the present one), in general the DBs can become
dynamical variable dependent \cite{11}
due to the $\{A,\Theta _\alpha \}$ and
$\Delta _{\alpha \beta }$
 terms, leading to problems for
the quantization programme. To cure this type of pathology, BT
formalism is a systematic framework where one introduces the BT
variables $\phi
 ^\alpha _a $, obeying
\begin{equation}
\{\phi ^\alpha ,\phi ^\beta \}=\omega ^{\alpha \beta}=
-\omega ^{\beta \alpha},
\label{bt}
\end{equation}
where $\omega ^{\alpha \beta}$ is a constant (or at most
 a c-number function) matrix, with the aim of modifying the SCC
$\Theta _\alpha  (q)$ to $\tilde \Theta _\alpha  (q,
\phi ^\alpha )$
such that,
\begin{equation}
\{\tilde\Theta _\alpha (q,\phi ) ,\tilde\Theta _\beta (q,\phi )\}=0
~~;~~\tilde\Theta _\alpha (q,\phi )=\Theta _\alpha (q)+
\Sigma _{n=1}^\infty \tilde\Theta ^{(n)} _\alpha (q,\phi )~~;~~
\tilde\Theta ^{(n)}\approx O(\phi ^n)
\label{b1}
\end{equation}
This means that $\tilde\Theta  _\alpha $ are now FCCs and in
particular abelian \cite{12}.
The explicit terms in the above expansion are \cite{12},
\begin{equation}
\tilde\Theta ^{(1)}_\alpha =X_{\alpha \beta }\phi ^\beta ~~;~~
\Delta _{\alpha \beta }+X_{\alpha \gamma }
\omega ^{\gamma \delta }X_{\beta \delta }=0
\label{b2}
\end{equation}

\begin{equation}
\tilde\Theta ^{(n+1)}_\alpha =-{1\over{n+2}}
\phi^{\delta }\omega _{\delta \gamma }X^{\gamma \beta }B^{(n)}_{\beta \alpha }~~;~~n\ge 1
\label{b3}
\end{equation}

\begin{equation}
B^{(1)}_{\beta \alpha }=
\{\tilde\Theta ^{(0)} _\beta ,\tilde\Theta ^{(1)} _\alpha \}_{(q)}-
\{\tilde\Theta ^{(0)} _\alpha ,\tilde\Theta ^{u(1)} _\beta \}_{(q)}
\label{b4}
\end{equation}

\begin{equation}
B^{(n)}_{\beta \alpha }=
\Sigma _{m=0}^n
\{\tilde\Theta ^{(n-m)} _\beta ,\tilde\Theta ^{(m)} _\alpha \}_{(q,p)}+
\Sigma _{m=0}^n
\{\tilde\Theta ^{(n-m)} _\beta ,\tilde\Theta ^{(m+2)} _\alpha \}_{(\phi )}
~~;~~n\ge 2
\label{b5}
\end{equation}
In the above, we have defined,
\begin{equation}
X_{\alpha \beta }X^{\beta \gamma }=
\omega _{\alpha \beta }\omega ^{\beta \gamma }
=\delta ^\gamma _\alpha \delta  .
\label{b6}
\end{equation}
A very useful
idea is to introduce the improved function $\tilde f(q)$ \cite{12}
 corresponding to each $f(q)$,
\begin{equation}
\tilde f(q,\phi )\equiv f(\tilde q)
=f(q)+\Sigma _{n=1}^\infty \tilde f(q,\phi )^{(n)}~~
;~~\tilde f^{(1)}=-
\phi^{\beta }\omega _{\beta \gamma }X^{\gamma \delta }\{
\Theta_\delta ,f\}_{(q)}
\label{b7}
\end{equation}

\begin{equation}
\tilde f^{(n+1)}=-{1\over{n+1}}
\phi^{\beta }\omega _{\beta \gamma }X
^{\gamma \delta }
G(f)^{\lambda (n)}_\delta ~~;~~n\ge 1
\label{b8}
\end{equation}

\begin{equation}
G(f)^{(n)}_{\beta }=
\Sigma _{m=0}^n
\{\tilde\Theta ^{ (n-m)} _\beta ,\tilde f^{(m)}\}_{(q)}+
\Sigma _{m=0}^{(n-2)}
\{\tilde\Theta ^{ (n-m)} _\beta ,\tilde f^{(m+2)}\}_{(\phi )}
+\{\tilde\Theta ^{ (n+1)} _\beta ,\tilde f^{(1)}\}_{(\phi )}
\label{b9}
\end{equation}
which have the property
$\{\tilde\Theta _\alpha (q,\phi ) ,\tilde f(q,\phi )\}=0$.
Thus the improved functions are FC or equivalently gauge invariant.
The subscript $(\phi )$ and $(q)$ in the PBs indicate the improved variables
with respect to which the PBs are to be taken.
It can be
proved that extensions of the original FCC $\Psi _l $ and Hamiltonian
 $H$ are simply,
\begin{equation}
\tilde \Psi _l=\Psi (\tilde q)~~;~~
\tilde H=H (\tilde q).
\label{b10}
\end{equation}
One can also reexpress the converted SCCs as
$\tilde\Theta ^\mu_\alpha \equiv \Theta ^\mu_\alpha (\tilde q)$.
The following identification theorem holds,
\begin{equation}
\{\tilde A,\tilde B \}=\tilde {\{A,B\}_{DB}}~~;~~
\{\tilde A,\tilde B \}\mid _{\phi =0}=\{A,B \}
 _{DB}~~;~~\tilde 0=0.
\label{b11}
\end{equation}
Hence the outcome
of the BT extension is the closed system of FCCs with the FC
Hamiltonian given below,
\begin{equation}
\{\tilde \Theta _\alpha ^\mu ,\tilde \Theta _\beta ^\nu\}=
\{\tilde \Theta _\alpha ^\mu ,\tilde \Psi _l\}=
\{\tilde \Theta _\alpha ^\mu ,\tilde H\}= 0~~;~~
\{\tilde \Psi _l ,\tilde \Psi _n\}\approx 0 ~;~
\{\tilde \Psi _l ,\tilde H\}\approx 0.
\label{b12}
\end{equation}
In general, due to the non-linearity in the SCCs, the extensions
 in the improved variables, (and subsequently in the FCCs
 and FC Hamiltonian), may
turn out to be infinite series. This type of situation has been encountered
 before \cite{11}. Fortunately this complication does not surface here. The are no FCCs in our system, (analogous to $\Psi _l$) and identifying the SCCs $\eta _\alpha $ in (\ref{10}) in the present case with $\Theta _\alpha $ in (\ref{01}), we compute the BT extended FCCs as,
  \begin{equation}
 \tilde \eta _1\equiv \eta _1+\phi _1~;~~\tilde \eta _2\equiv \eta _2
 -\Pi _\theta \phi _2~,~\{\tilde {\eta _1},\tilde {\eta _2}\}=0.
 \label{11}
 \end{equation}
 The BT fields obey $\{\phi _1(x),\phi _2(y)\}=\delta (x-y)$.
  To ensure that there are no further constraints,
  we need a Hamiltonian that commutes with the FCCs.
The following variables,
\begin{equation}
\tilde \theta =\theta +\alpha \phi _2 ~,~\tilde \Pi _\theta =\Pi _\theta ~
,~\tilde \alpha =\alpha +{{\phi _1}\over {\Pi _\theta }} ~,~
\tilde \Pi _\alpha = \Pi _\alpha - \Pi _\theta \phi _2 ~,~\tilde \beta =\beta -\phi _2 ~,
~\tilde \Pi _\beta =\Pi _\beta ~,~\tilde \phi _i =0 .
 \label{12}
 \end{equation}
are gauge
  invariant \cite{12} in the sense that they commute with the FCCs.
   Hence {\it all} quantities written in terms of the redefined variables
   are gauge invariant in the extended
  space. In particular, the modified (free) Hamiltonian reads
   \begin{equation}
 \tilde {\cal H}\mid _{free}={1\over 2}(\tilde \Pi _\theta \tilde {v_i}\tilde {v_i})
 ={1\over 2}\Pi _\theta [\partial _i(\theta +\alpha \phi _2)
 +(\alpha +{{\phi _1}\over {\Pi _\theta }})\partial _i(\beta -\phi _2)]^2.
 \label{13}
 \end{equation}
The remaining interaction terms in $H$, if present,
 will also be extended in a similar way.
 This Hamiltonian (\ref {13}) together with the FCCs (\ref{11})
and the canonical phase space is the
 gauge invariant system we were looking for. This constitutes one of the major
results of the paper.

It may be worthwhile at this point to comment on the apparently
involved structure of the extended Hamiltonian (\ref{13}) with a
dynamical variable residing in the denominator. As has been
emphasized in the Introduction, this complexity is unavoidable.
However, even then, this extended system together with its
canonical phase space is more suitable for quantization in a
perturbative framework. For example, for "small" or "large" $\Pi
_\theta $, {\it i.e.} density, in some scale, one can expand $\Pi
_\theta $ around a background or expand in terms of $(1- \Pi
_\theta )$ in the offending term in (\ref{13}).

It will be appropriate to make a comparison between \cite{new0} and our analysis. Essentially the aim of Hamiltonian formulation of a dynamical system is to provide a Hamiltonian function and a set of PBs, which will reproduce (via the Hamiltons equations of motion) a {\it given} set of equations of motion of the system. Indeed, there are some consistency conditions for the PBs (such as Jacobi identity etc.). Now, the basic set of dynamical variables in \cite{new0} are different from our set, although it is obvious that there exists a mapping between the two sets, once the additional constraint of constant density is imposed on our model. But more importantly, it should be noted that the basic quantum commutation relations in \cite{new0} are operator valued and hence perturbative computations will be difficult to perform. On the other hand, the present system is embedded in the BT extended space, which is completly canonical ({{\it i.e.} not operator valued) by construction !
and thus is amenable to conventional perturbative analysis. Because of the introduction of auxiliary fields a standard BRST \cite{brst} quantization is to be performed, which has not been carried through here.

To make contact with the physical system, another alternative is to reduce
the dimension of the BT extended
phase space 
by additional gauge fixing constraints, (two in this case,
 $\tilde {\eta _3}$ and $\tilde {\eta _4}$,
 corresponding to two FCCs), with the
 only restriction that $\tilde {\eta _a},~a=1,..,4$ constitute an
 SCC system that is $det \mid \{\tilde \eta _a,\tilde \eta _b\}\mid \ne 0$. A
 consistency check is to see that
 the original system is recovered in
 the so called unitary gauge, $\tilde \eta _3\equiv \phi _1\approx 0~~,~~
\tilde \eta _4 \equiv \phi _2\approx 0$. It is interesting to note that in the extended space, the Hamiltonian or any other observable can, in general, depend upon the Clebsch variables, which are not expressible in terms of the velocity $v_i$. However, it should be remembered that they describe physically allowed systems since they are gauge equivalent to the physical system.

It might be convenient, (although not necessary), to consider the gauges
of the form
 $ \tilde {\eta _3}\equiv \phi _1-F~,~\tilde {\eta _4}\equiv \phi _2
-G$, to remove  the BT fields directly. $F$ and $G$  can contain
the physical fields as well.
For a particular gauge, one has
 to construct the corresponding $DB$s and compute the equations
 of motion using the $DB$s in reduced phase space, where the SCCs have
been used strongly. Once again, the degrees of freedom count agrees
with the original one.
Consider the special class of gauge transformations: $\phi _1=0~,~
\phi _2=constant $. These will {\it not} change the $(v_i,\rho )$
 algebra. Hence they can be identified as the conventional canonical
  transformations.
Furthermore, additional constraints, such as incompressibility
\cite{8}, can
be included in this setup in the form $\rho =constant$, which under
time persistence generates another constraint $\partial _i\tilde
{v_i}$. This SCC pair leads to \cite{8}.

The constants of motion for the free theory are obviously the energy
$\tilde H$, the momenta $\tilde P_i=\int (\rho \partial _i\theta +
\Pi _\alpha \partial _i\alpha + \Pi _\beta \partial _i\beta +
\phi _2 \partial _i\phi _1 )$, the angular momenta
$\tilde L^{ij}=\int (r^i\tilde {\cal P}^j-r^j\tilde {\cal P}^i)$ and the boost generator
$\tilde B^i=t\tilde P_i-\int (r_i\rho )$, effecting the
transformation
$$\{\tilde {v_i},u_j\tilde B_j\}=-t(u_j\partial _j)\tilde v_i+u_i~,~
\{\rho,u_j\tilde B_j\}=-t(u_j\partial _j)\rho . $$

Obtaining the Lagrangian is indeed straightforward. The first order
 form is
 $${\cal L}=\Pi _\theta \dot \theta +\Pi _\alpha \dot \alpha
+\Pi _\beta \dot \beta  +\phi _2 \dot \phi _1 -\tilde {\cal H }
-\lambda _1\tilde {\eta _1}  -\lambda _2\tilde {\eta _2}
$$
\begin{equation}
\equiv \Pi _\theta \dot \theta  +\phi _2 \dot \phi _1  +\dot \beta
(\alpha \Pi_\theta +\phi _1)+\dot \alpha \Pi_\theta \phi _2
-\tilde {\cal H}
-\lambda _1\tilde {\eta _1}  -\lambda _2\tilde {\eta _2},
\label{l}
\end{equation}
where $\lambda _1$ and $\lambda _2 $ are multiplier fields and
some of the variables have been removed using the equations of
motion. The generic gauge transformation is defined as
\begin{equation}
\delta A=\{\int (\epsilon \Psi ),A\},
\label{gtr}
\end{equation}
where $A$, $\epsilon $ and $\Psi $ represent some operator,
infinitesimal gauge transformation parameter and an FCC
respectively. At this stage, one can check explicitly that
(\ref{l}) is invariant under the following two independent sets of
gauge transformations corresponding to the two FCCs,
$$
\tilde {\eta _1}\rightarrow \delta _1\Pi _\theta =0~,~\delta _1\theta =
-\alpha \psi _1~,~\delta _1 \beta =\psi _1,\delta _1 \alpha =0~,~
\delta _1 \phi _1
=0~,~\delta _1 \phi _2=\psi _1~;$$
\begin{equation}
\tilde {\eta _2}\rightarrow \delta _2\Pi _\theta =0~,~\delta _2\theta =
\phi _2 \psi _2~,~\delta _2 \beta =0,\delta _2 \alpha =
-\psi _2~,~\delta _2 \phi _1
=\Pi _\theta \psi _2~,~\delta _2 \phi _2=0,
\label{gt}
\end{equation}
where $\psi _1$ and $\psi _2$ are gauge transformation parameter functions.
Naively taking the unitary gauge, {\it i.e.}  $\phi _1=\phi _2 =0$, we can
recover the Lagrangian posited in \cite{1}.

We now discuss briefly the relativistic generalization of the parent
free theory.
In the relativistic generalization of the free theory \cite{1},
the Lagrangian is expressed as
$$
{\cal L}_{rel}=j^\mu a_\mu -(j^\mu j_\mu )^{{1\over 2}},
$$
\begin{equation}
a_\mu =\partial _\mu \theta +\alpha \partial _\mu \beta ~,~
j^\mu =(\rho ,~\rho v^i) .
\label{k}
\end{equation}
Notice that the symplectic structure does not change from the
non-relativistic one.
Expansion of the square root as
$$\rho (1+v^iv_i)^{{1\over 2}}\approx \rho (1+{1\over 2}v^iv_i +...),$$
and dropping the uninteresting $\int \rho $ term
\cite{1}, (since it can only
influence the time evolution of $\theta $ by a constant translation),
 we can recover the non-relativistic Lagrangian in (\ref{g1})
with $U=0$.
The   Hamiltonian now is modified to
\begin{equation}
{\cal H}_{rel}=\rho [v_iv_i+(1-v_iv_i)^{{1\over 2}}],
\label{l1}
\end{equation}
which changes the equations of motion to the following:
$$\dot\alpha =L_i\partial _i\alpha ~~;~~\dot\beta =L_i\partial _i\beta ~~;~~
\dot\theta =-L_i\alpha \partial _i\beta +
[v_iv_i+(1-v_iv_i)^{{1\over 2}}]
~~;~~\dot\rho=\partial _i(\rho L_i),$$
\begin{equation}
L_i={1\over \rho }{{\partial {\cal H}_{rel}}\over {\partial v_i}}
=v_i[2-(1-v_iv_i)^{-{1\over 2}}].
\label{m}
\end{equation}
Notice that in the lowest order, $L_i\approx v_i+O(v^3)$, the previous
equations are recovered.

BT extension of the relativistic model is straightforward since the
symplectic structure remains unchanged from the non-relativistic one. One
only has to replace the original variables by their gauge invariant
 counterpart (\ref{12}) in the covariant expressions (\ref{k}) and
(\ref{l1}).

We briefly remark on the canonical quantization of the BT extended fluid model. 
Indeed, the BT extension has rendered the phase space canonical and the classical variables are simply elevated to quantum operators with the generic equal time commutators being,
$$[\phi (x)~,~\pi _\phi (y) ]=i\hbar \delta (x-y).$$
The FCCs (\ref{11},\ref{10}) are taken in to consideration by invoking the Dirac quantization prescription \cite{10}, in which the physical states are defined as \cite{1sg,2sg}
\begin{equation}
\mid Ph~State>\equiv \delta (\tilde \eta _1)\delta (\tilde \eta _2)\mid State>~,
\label{phy}
\end{equation}
indicating that the physical states are annihilated by the FCCs. One can immedietly get some idea on the qualitative nature of the (lowest order) quantum corrections involved, say in the energy spectrum, by sandwitching the extended space FC Hamiltonian operator (\ref{13}) between the physical states as defined above. Following \cite{1sg,2sg} we can resort to Weyl ordering the quantum operator products and for simplicity let us exploit the unitary gauge. Naively it might seem that there can appear non-trivial quantum corrections. However it is straightforward to check that the structure of the constraints in the unitary gauge and the derivative  operators involved (in the Hamiltonian) conspire to produce the quantum corrections in the form of total derivatives only, which are assumed to vanish. Notice that in an apparantly similar circumstance in \cite{2sg}, in the $CP^1$-model with Hopf interaction, non-zero quantum corrections appeared in the form of total derivatives in t!
he topologically non-trivial sector only. No such topological protection is present here.
\section{Reduced space quantization}
\label{sec:ReducedSpaceQuantization}
As mentioned in the Introduction, we now discuss briefly the alternative
scheme \cite{vy} of inducing gauge invariance in a system subjected to
SCCs only, {\it without} enlarging the phase space, that is no BT fields
are introduced. This prescription is particularly suitable for the special
 case of only two SCCs, as is the case here, (\ref{10}). The primary idea is
to consider (a suitably modified form of) one of the SCCs as the only FCC and
discard the other SCC, which may be thought of as a particular gauge fixing condition
 one is free ignore. There is a prescription \cite{vy} by which gauge invariant
counterparts of all the variables, (and hence the Hamiltonian
 and other relevant quantities as well), can be constructed. In fact, this formalism
and the BT extended scheme can be related in a formal way, at least for two SCCs \cite{vy}, which
however will not be discussed here. For two generic SCCs $Q_1$ and $Q_2$, with
$\{Q_1,Q_2\}=\Lambda\ne 0$, the SCC pair can be replaced by a single FCC
 $\chi =\Lambda ^{-1}Q_1$ with $Q_2$ dropped. Obviously taking $Q_2$ instead of
$Q_1$ in $\chi $ is also a valid choice. It can be checked that for any degree of
freedom $A$, $\{\chi ,\tilde A\}=0$, where $\tilde A $, defined below, is the gauge
invariant counterpart of $A$,
\begin{equation}
\tilde A=A-Q_2\{\chi ,A\}+{1\over {2!}}Q_2^2\{\chi ,\{\chi ,A\}\}-
{1\over {3!}}Q_2^3\{\chi ,\{\chi ,\{\chi ,A\}\}\}+...
\label{v1}
\end{equation}
In the present case, with the SCC system (\ref{10}) the gauge invariant fluid models are\\
Case I:\\
$$\chi ={{\Pi _\alpha }\over {\Pi _\theta }},$$
\begin{equation}
\tilde \theta =\theta +
(\alpha \Pi _\theta -\Pi _\beta ) {{\Pi _\alpha }\over {\Pi _\theta ^2 }},~~~
\tilde \alpha ={{\Pi _\beta }\over {\Pi _\theta }}.
\label{v2}
\end{equation}
The gauge variation of the first order Lagrangian also vanishes,
$$
\delta {\cal L}=\{\int d^3x (\lambda \chi ),(\Pi _\theta \dot \theta +\Pi _\alpha \dot \alpha
+\Pi _\beta  \dot \beta )-\tilde H \}$$
$$=\lambda \dot \chi \approx 0.$$
Case II:\\
$$
\chi =\alpha -{{\Pi _\beta }\over {\Pi _\theta }}$$
\begin{equation}
\tilde \theta =\theta +{{\Pi _\alpha \Pi _\beta }\over {\Pi _\theta ^2}} ,~~~
\tilde \beta =\beta -{{\Pi _\alpha }\over {\Pi _\theta }}.
\label{v3}
\end{equation}
The gauge variation of the first order Lagrangian again vanishes,
$$\delta {\cal L}=\lambda \dot \chi \approx 0.$$

Let us now make a direct contact between the gauge invariant theories formulated in the BT
extended scheme in section {\bf 3} and in the present section. Obviously, the former is the
more general one as it contains two FCCs whereas the latter has a single FCC. Considering
the FCC of case I above in the reduced space scheme, the gauge transformations are,
\begin{equation}
\delta \alpha =-\frac{\epsilon}{\Pi _\theta }~~,~~\delta \theta =
\epsilon \frac{\Pi _\alpha}{\Pi _\theta ^2}.
\label{red}
\end{equation}
Now in the BT formulation, choose a gauge fixing condition as
\begin{equation}
{\cal G}\equiv \phi _2-\frac{\Pi _\alpha}{\Pi _\theta ^2}.
\label{gch}
\end{equation}
This gauge will render the pair $({\cal G},\tilde \eta _1 )$ second class while keeping the first class nature of $\tilde \eta _2$ intact. Presently, computing DBs with respect to the above SCC pair and using them to compute the gauge variations, one finds that on the FCC surface $\tilde \eta _2\approx 0$ the gauge transformations in the BT and reduced schemes are identical. For a different gauge condition the FCC in case II above in the reduced scheme can also be matched to the BT results.
This verification is the analogue of \cite{wot} for a general fluid.
This concludes our analysis of the quantization problem of the Hamiltonian fluid model both from an extended space and reduced space points of view. The hidden gauge invariance
of the fluid system has also been revealed in both the BT and reduced schemes. Indeed, the above correspondence between the results obtained in the two distinct methods is important as it connects the gauge invariance in the enlarged (BT) space to the one in physical space.

\section{Discussions}
\label{sec:Discussions}
The Hamiltonian formulation of isentropic, inviscid fluid in three spatial dimensions has been studied as a prelude to its quantization. The Poisson bracket structure, yielding the fluid equations of motion, is non-canonical (in particular field dependent). This creates problem for the convenional quantization programme since the quantum commutators (via the correspondence principle) will also aquire a field dependence and as a result even the propagators can not be defined properly.

This motivates us to exploit the Batalin-Tyutin quantization
scheme \cite{12} in which the constraints are taken in account in
a canonical extended phase space. The system is also converted to
a gauge theory. A second method, the reduced space scheme
\cite{vy} is also applied to the fluid model. Here also the fluid
model is converted to a gauge theory but contrary to the previous
framework \cite{12}, the phase space extension is not required.
This hidden gauge invariance in the fluid model in the latter
formulation has also been explicitly connected to the former one.
This hidden gauge symmetry actually corresponds to the gauge
invariance present in the equivalent relativistic membrane theory.
This is a generalization of the works of \cite{wot, 10a}.

The presence of gauge invariance offers more freedom in the
analysis of a theory and apparently different models can become
identified as gauge equivalent ones, so that results obtained in
one model can be carried to the other one. For instance, generally
one considers the fluid system, in physical situation, as being
subjected to a pressure term $({{\partial _i
 p(x)}\over {\rho }})$ and a constant force such as gravity. By a
suitable choice of (translation symmetry breaking) gauge, one can
generate these terms in the Euler equation. However, even in the
linear approximation, there will be additional terms in the Euler
equation besides the above ones and the continuity equation will
also be modified. One can say that this set is gauge equivalent to
the free theory since the latter is reproduced in the unitary
gauge. Indeed. one can trade one type of interaction and source
with another one by exploiting the BT gauge equivalence and one
set might be better suited to simulate experimentally or analyze
theoretically.

The Batalin-Tyutin construction is primarily aimed at providing a
canonical framework for quantizing the fluid system. It will indeed
be interesting to study the quantized version of the above model presented here in more detail.

Furthermore, gauge invariant interactions of the form $\epsilon
_{\mu\nu\sigma\lambda}F^{\mu\nu}\theta \partial ^\sigma \alpha
\partial ^\lambda \beta $ can also be introduced in the action
which will alter the symplectic structure and the resulting theory
can be studied perturbativly.

\vskip .2cm
{\it Acknowledgements}: I thank Dr. K. Kumar for numerous stimulating discussions.


\newpage

\begin{thebibliography}{99}
\bibitem{br} For a review see A.Giveon and D.Kutasov, Rev.Mod.Phys. 71 983(1999).
\bibitem{br1} J.Goldstone, (unpublished);  M.Bordemann and J.Hoppe, Phys.Lett. B317 315(1993).
\bibitem{1} For a review, see R.Jackiw, physics/0010042.
\bibitem{2}
 R.Jackiw and A.P.Polychronakos, HEP-TH/9809123;
D.Bazeia and R.Jackiw, Ann.Phys.(NY), (HEP-TH/9803165); R.Jackiw
and A.P.Polycronakos, Comm.Math.Phys. 207 107(1999),
 D.Bazia, Phys.Rev. D59 085007(1999).
 \bibitem{lan}I.Landau and E.Lifshitz, {\it Fluid Mechanics}, (2nd ed., Pergamon,
 Oxford, UK, 1987); P.Olver and Y.Nutku, J.Math.Phys. 29 1610(1988).
 \bibitem{bar} B.Barbishov and N.Chernikov, Sov.Phys. JETP 24 1371(1967).
 \bibitem {landau}L.D.Landau, ZETF 11 592(1941) [English translation: J.Phys.USSR 5 71(1941)].
 \bibitem{5} H.Lamb, {\it Hydrodynamics}, (Cambridge University Press,
Cambridge, UK, 1932);
For subtleties involved in Clebsch parametrization, see
S.Deser, R.Jackiw and A.Polychronakos, Phys.Lett.A 279 151(2001).
\bibitem{wot}C.Neves and C.Wotzasek, HEP-TH/0105281.
 \bibitem{10} P.A.M.Dirac, {\it Lectures on Quantum Mechanics}, (Yeshiva
University Press, New York, 1964).
\bibitem{12} I.A.Batalin and I.V.Tyutin, Int.J.Mod.Phys. A6 3255(1992).
\bibitem{10a} C.P.Natividade and H.Boschi-Filho, Phys.Rev.D62 025016
(2000).
\bibitem{vy} P.Mitra and R.Rajaraman, Ann.Phys. (NY) 203 157(1990);
R.Anishetty and A.S.Vytheeswaran, J.Phys.A 26, 5613(1993).
\bibitem{3} A.Thellung, Physica 29 227(2953).
\bibitem{4} R.Salmon, Ann.Rev.Fluid Mech. 20 225(1988).
\bibitem{st}S.Bergliaffa, K.Hibbard, M.Stone and M.Vissar, arXiv:COND-MAT/0106255.
\bibitem{new0}M.Rasetti and T.Regge, Physica 80A 217(1975).
\bibitem{baz}See the papers by Bazeia and Jackiw in \cite{2}.
\bibitem{mad}E.Madelung, Z.Phys. 40 322(1926); E.Merzbacher, {\it Quantum Mechanics},
(3rd ed., Wiley, New York, 1998).
\bibitem{jev}A.Jevicki, Phys.Rev. D57 5955(1998).
\bibitem{sh} A.M.J.Shakel. Mod.Phys.Lett.B 4 927(1990).
\bibitem{kam}A.Yu.Kamenshchik, U.Moschella and V.Pasquier, Phys.Lett.B 487 7(2000).
\bibitem {new1} I.E.Dzyaloshinskii and G.E.Volovick, Ann.Phys. 125 67(1980); G.E.Volovick and V.S.Dotsenko Jr., JETP Lett. 29 576(1979); A.F.Andreev and M.Yu.Kagan, Zh.Eksp.Teor.Fiz. 86 546(1984) [Sov.Phys. JETP 59 318(1984)]; G.E.Volovick, JETP Lett. 64 845(1996).
\bibitem{8} S.Nguyen and L.A.Tursky, Physica A290 431; {\it ibid} A272 48(1999).
{\it ibid} 57 432(1972); S.Ghosh, arXiv HEP-TH/0106166.
\bibitem {new2} L.Faddeev and R.Jackiw, Phys.Rev.Lett. 60 1692(1988).
\bibitem {11}See for example N.Banerjee, R.Banerjee and S.Ghosh, Nucl.Phys.
B427 257(1994).
\bibitem{brst}C.Becchi, A.Rouet and R.Stora, Phys.Lett. 52B 344(1979); I.V.Tyutin, Lebedev Preprint FIAN, No. 39 (1975).
\bibitem{1sg}W.Oliveira and J.A.Neto, Int.J.Mod.Phys. A12 4895(1997).
\bibitem{2sg} B.Chakraborty, S.Ghosh and P.P.Malik, Nucl.Phys. B600 351(2001).
\end{thebibliography}
\end{document}